\documentclass{vgtc}                          

\let\ifpdf\relax
\usepackage{graphicx}
\usepackage{times}
\usepackage{mathptmx}
\usepackage{color}
\usepackage{textcomp}
\usepackage{booktabs}
\usepackage{enumitem}
\usepackage{ccicons}
\usepackage{todonotes}
\usepackage[hang,footnotesize]{subfigure}

\usepackage{booktabs}
\usepackage{tabu}

\usepackage{microtype}

\usepackage[pdftex,final]{hyperref}

\usepackage{tabularx}

\graphicspath{{figures/}{pictures/}{./}}
\DeclareGraphicsExtensions{.pdf,.png,.jpg,.jpeg}

\hypersetup{
  pdftitle={},
  pdfsubject={},
  pdfauthor={},
  pdfcreator={LaTeX2e},
  pdfkeywords={},
  plainpages=false,       
  hypertexnames=false,    
  bookmarks=true,         
  bookmarksnumbered=false,
  bookmarksopen=true,     
  bookmarksopenlevel=3,   
  pdfpagemode=UseNone,    
  pdfstartview=Fit,
  pdfborder={0 0 0},      
  breaklinks=true,        
  colorlinks=false,       
  nesting=true}

\ifx\pdfoutput\undefined%
  \renewcommand{\pdfbookmark}[3][]{}
\fi%

\newcommand{\eg}{e.\,g.}
\newcommand{\ie}{i.\,e.}

\onlineid{0}

\vgtccategory{Research}

\vgtcinsertpkg

\title{Usability Comparison of Mouse, Touch and Tangible Inputs\\[.5ex] for 3D Data Manipulation}

\author{Lonni Besan\c{c}on\thanks{e-mail: \{lonni.besancon,tobias.isenberg\}@inria.fr}\\ %
        \parbox{1.4in}{\scriptsize \centering INRIA Saclay \\ Univ. Paris-Saclay} %
\and Paul Issartel\thanks{e-mail: \{paul.issartel,mehdi.ammi\}@limsi.fr}\\ %
     \parbox{1.4in}{\scriptsize \centering LIMSI-CNRS\\ Univ. Paris-Saclay} %
\and Mehdi Ammi\footnotemark[2]\\ %
     \parbox{1.4in}{\scriptsize \centering LIMSI-CNRS \\ Univ. Paris-Saclay}%
\and Tobias Isenberg\footnotemark[1]\\ %
     \parbox{1.4in}{\scriptsize \centering INRIA Saclay}}

\abstract{

We evaluate the performance and usability of mouse-based, touch-based, and tangible interaction for manipulating objects in a 3D virtual environment. This comparison is a step toward a better understanding of the limitations and benefits of these existing interaction techniques, with the ultimate goal of facilitating the integration of different 3D data exploration environments into a single interaction continuum. For this purpose we analyze participants' performance in 3D manipulation using a docking task. We measured completion times, docking precision, as well as subjective criteria such as fatigue, workload, and preference. Our results show that the three input modalities provide similar levels of precision but require different interaction times. We also discuss our qualitative observations as well as people's preferences and put our findings into context of the practical application domain of 3D data analysis environments.

} 

\CCScatlist{ 
  \CCScat{H.5.2}{Information Systems}{User Interfaces}{Input devices and strategies}
}

\renewcommand{\marginpar}[1]{}

\begin{document}


\firstsection{Introduction}

\maketitle

Many application domains rely on effective, efficient, and intuitive means of interacting with 3D data \cite{Keefe:2010:IVI,Munzner2006}. Traditionally, this interaction has often relied on mice and keyboards as the prime means of providing input. Recent developments of interaction technology, however, have led to new input modalities becoming available, in particular touch-based input \cite{Shneiderman:1991:TSN,Wigdor:2011:BNW} and tangible interaction \cite{Ishii:2008,Shaer2010}. Several researchers as well as tool developers have thus started to explore the use of these novel input modalities for the interaction with 3D data. Nevertheless, the three input modalities---mouse, touch, and tangibles---are not identical in characteristics such as their capabilities or usability: they all have advantages and disadvantages depending on the given interaction goal and the respective application domain. For example, while for a game we may expect a tangible input device to be easy and intuitive to use, scientific visualization applications may require a level of precision that one could expect to better be provided by touch-based input or, in particular, mouse and keyboard interfaces.

We are convinced that, in the future, no means of interaction will prevail over the others. It is crucial, therefore, to better understand the characteristics and differences between the three modalities. We see our work as the first step toward the realization of an interaction continuum for 3D data analysis in which researchers can seamlessly transition from one interaction scenario that relies on one technique to an other scenario---depending on the task at hand. We thus have conducted a comparative study of the mentioned three input modalities for 3D interaction. The study consists of 15 docking tasks for each of the three modalities. During the study, we measured the participants' precision (\ie, rotational difference and Euclidean distance), their perceived fatigue levels, and their perceived workload. We also took into account other factors such as participants' preferences and their general feedback for each technique.

The study showed that mouse, touch, and tangible input are valid means for controlling 3D manipulations. Interestingly, we found that all three input modalities allow users to achieve the same level of precision. In contrast to the precision results, the study revealed differences with respect to task completion times and preferences. Qualitative observations of the participants during the study provided insights on what users tend to do when facing a docking task with these three input techniques.

In summary, our contributions comprise
\begin{itemize}[itemsep=.25\lineskip,topsep=3\lineskip]
\item an in-depth analysis of people's understanding and usage of the mouse, touch, and tangible input modalities for 3D interaction,
\item a study design that compares the three methodologies without placing a focus on a single evaluation parameter, and
\item the discussion of qualitative observations and people's preferences in the context of 3D data analysis environments.
\end{itemize}
Our findings provide insights on the advantages and disadvantages of the techniques and serve as a basis for their further development and evaluation, in particular for 3D visualization.

\section{Related Work}

Much of past work has focused on the comparison of interaction techniques or devices. In many cases, academic studies compare novel technique(s) or device(s) to established ones. For instance, many studies were conducted to compare the advantages and limitations of mouse interaction compared to touch interactions for tasks as various as selection, pointing, exploration etc. (\eg, \cite{Forlines2007,Kin2009,Sasangohar2009}). Our review of the literature, however, revealed a lack of studies that would analyze these modalities for 3D manipulation tasks---only few researchers actually conducted such analyses \cite{Chen1988,Hinckley1997,Tuddenham2010,Yu:2010:FDT}.

Among these, Chen et al.\ \cite{Chen1988}---as early as in the 1980s---and later Hinckley et al.\ \cite{Hinckley1997} compared input techniques for 3D manipulation. Both studies, however, narrowly focused on rotation and did not take into account other parameters such as Euclidean distance to the target or usability. Tuddenham et al. \cite{Tuddenham2010} compared mouse, touch, and tangible interaction for a matching task which was constrained on a tabletop, thus constraining the interaction to two dimensions. They measured the time required to complete the task, the ease of use, and people's preference. Yu et al.\ \cite{Yu:2010:FDT}, finally, compared mouse and touch interaction to validate their FI3D widget for 7DOF data navigation. In our work, in contrast, we aim to get a more holistic and general view of how the different input modalities affect the interaction with 3D shapes or scenes, ultimately to better understand how they can support the analysis of complex 3D datasets.

Most of the comparative studies also focus on comparing either mouse and touch interaction techniques or tangible and touch (and many of these do that for 2D tasks). The literature indeed contains many papers comparing touch and mouse input for a whole variety of tasks and a whole variety of parameters: speed \cite{Forlines2007,Sears1991}, error rate \cite{Forlines2007,Sears1991}, minimum target size \cite{Albinsson2003}, etc. Similarly, much research has compared touch-based with tangible interaction for tasks as various as puzzle solving \cite{Terrenghi2007,Wang:2008:CTM}, layout-creation \cite{Lucchi2010}, photo-sorting \cite{Terrenghi2007}, selecting/pointing \cite{Raynal2010}, and tracking \cite{Jansen2012}. Most of the work comparing tangible interfaces to other interfaces builds on the assumption that physical interfaces, because they mimick the real world, are necessarily better. However, this assumption was rightfully questioned by Terrenghi et al.\ \cite{Terrenghi2007}. A 2DOF input device such as a mouse may, in fact, perform well in a 3D manipulation task due to its inherent precision or people's familiarity with it. To better understand advantages and challenges of the three mentioned input modalities, we thus compare them with each other in a single study.

Esteves and Oakley \cite{Esteves2011} also emphasize the fact that most studies comparing tangible interaction to other interaction paradigms are hard to generalize due to the highly simplistic tasks asked of participants. Studies can thus only support very general claims on tangible interactions and its possible benefits. The lack of generalizability of such studies may also be explained by the overly focused participant groups in such studies. Very young participants often seem to be chosen to evaluate tangible interaction: for example, school-aged children were asked to evaluate the entertainment of Tangible User Interfaces (TUIs) \cite{Xie2008}, asked to solve puzzles \cite{Antle2009}, or asked to ``work together'' (collaboration tasks) to understand which paradigm can be used to reduce conflicts in collaboration tasks \cite{Marshall2009,Olson2011}. Similarly, Lucchi et al.\ \cite{Lucchi2010} asked college students to recreate layouts using touch and tangible interfaces. The learning effects of tangible interaction was also tested on non-adult participants in a study conducted by Price et al.\ \cite{Price2003}. We try to avoid this lack of generalizability by having a variety of participants and by using a task that is highly generalizable to 3D manipulation. We argue that this can be achieved by using a 3D docking task, which has often been used in the literature to evaluate new 6DOF devices \cite{Froehlich2006,Zhai1998}, new interaction techniques \cite{Hancock2007}, and for paradigm comparison studies \cite{Tuddenham2010} (for the latter, however, the docking was only conducted in two dimensions). We argue that using a low-level task such as 3D docking is the key to be able to generalize results from comparative studies.

Also related to our work is the topic of remote 3D manipulation through touch input that benefits from the increasing availability of large displays and the pervasive nature of mobile touch-enabled devices. For instance, Liang et al.\ \cite{Liang:2013:ISI} investigated the use of two back-to-back mobile devices---to facilitate touch input above and under the mobile device---with a combination of touch gestures and sensors to support rotation, translation, stretching, place slicing, and selection of 3D shapes. They also conducted an experiment to examine the use of dedicated regions on the mobile device to control objects or the 3D environment. Similarly, Du et al.\ \cite{Du:2011:TTM} investigated the use of a smartphone to navigate within a virtual environment on screen, while Katzakis et al.\ \cite{Katzakis:2012:PC3} examined the combination of mobile sensors and touch input for 3D translation and rotation through a docking task. Coffey et al.\ \cite{Coffey:2012:ISW}, however, used ``indirect'' touch manipulation to navigate and examine a volumetric dataset to overcome the inherent issues of touch interaction with stereoscopic rendering \cite{Valkov:2011:TSO}. We are interested, in contrast, in a more ``direct'' interaction\footnote{The terms ``direct'' and ``indirect'' interaction have to be used carefully. While mouse input is arguably indirect, tangible and touch input have both direct and indirect properties. Touch input, in our case, occurs directly on the displayed data (albeit on a projection of the real 3D shape) and is thus typically considered to be a direct interaction \cite{Knoedel2011,Levesque2011,Poupyrev:2003:TIS,Simeone:2015:CID}. Tangible input directly manipulates a 3D shape (tangible) where the virtual shape is thought to be, but our visuals are projected onto the separate display. We thus argue that touch and tangible interaction are more direct than mouse interaction.} which also displays the 3D information---we do not focus on remote manipulation by means of separate displays.

Our study mainly builds on the work by Hinckley et al.\ \cite{Hinckley1997} and Tuddenham et al.\ \cite{Tuddenham2010}. Hinckley et al.\ \cite{Hinckley1997} conducted comparative 3D docking studies focused on rotation with four different techniques including a 3D ball (our equivalent is a tangible interface) and a mouse. We go beyond their approach in that we consider a full 6\,DOF manipulation and evaluate more than time and accuracy. We go beyond Tuddenham et al.'s \cite{Tuddenham2010} approach in that we, while also comparing mouse, touch input, and tangible interfaces, use true 3D manipulation tasks---including for the tangible input device.

\section{Comparative Study}

As we aim to understand the use of mouse, touch, and tangibles for the manipulation of 3D scenes or datasets, our study investigates relevant tasks, in a realistic scenario, using a wide range of participants. Beyond time and error metrics, we observed people's actions, learnt about their realistic preferences, and their subjective ratings of the techniques. We aimed to understand four of Nielsen's \cite{Nielsen:1993:UE} five factors of usability: effectiveness, efficiency, satisfaction, error tolerance, and ease of learning. Error tolerance, was not within the scope of our study. The effectiveness is reflected by a precision score (in both angular and Euclidean distance), the efficiency by means of the time to complete the task, the satisfaction by looking at participants' answers to our questions, and the ease of learning by looking at the evolution of task completion times. 

\subsection{Task}

The 3D docking task we employ comprises translation in 3\,DOF, re-orientation in 3\,DOF, and precise final positioning of 3D shapes---actions similar to those that are common in and representative of interactive 3D data exploration.
A docking task (see other examples of docking task studies; \eg,  \cite{Chen1988,Froehlich2006,Hancock2007,Hinckley1997,Zhai1998}) consists of bringing a virtual object to a target position with a target orientation. The docking target is shown on the screen as a wire-frame version of the manipulated object. Such a docking interaction thus mimics many aspects of typical 3D interaction, even though an actual docking target may only implicitly exist in real-life scenarios.

\begin{figure}[tb]
  \centering
  \includegraphics[width=\columnwidth]{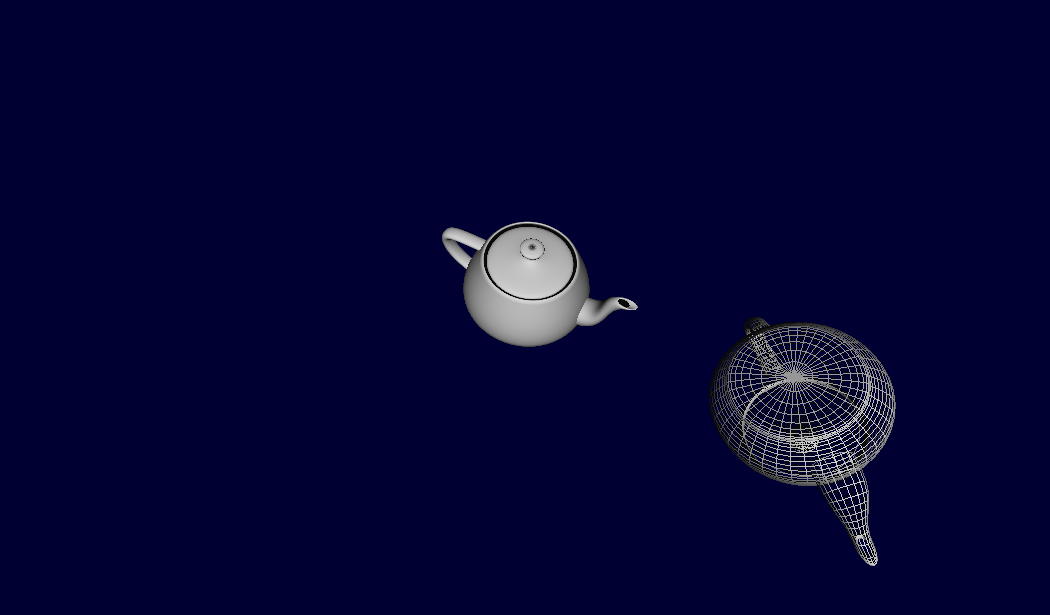}
	\caption{Screenshot of the task: participants were asked to move and orient the shaded object such that it matches the target (wireframe).}
	\label{fig:screenshot}
\end{figure}

In practice, we used the Utah teapot as the 3D object to manipulate as it is a generic shape most people understand and does not present any ambiguity in its orientation. We randomly generated and validated the starting and target positions beforehand (to ensure that all targets are reachable by all input modalities), yielding a pool of 15 valid starting and associated target positions (see \autoref{fig:screenshot}). Per input modality, we thus asked our participants to carry out 15 repetitions. For each of them we randomly selected the positions from the remaining positions in the pool. We used the same pool of positions for all input modalities. We counter-balanced the order of input modalities each participant saw to avoid a bias from learning effects. Our within-participants design thus comprised of 3 input modalities \texttimes\ 1 task \texttimes\ 15 trials = 45 trials in total for each participant.

Each trial was started and validated on a key press by the participant (similar to Chen et al.\ \cite{Chen1988} or Hinckley et al.\ \cite{Hinckley1997}). We considered using a pedal for validation (\eg, \cite{Hinckley1997}) but our pilots showed its triggering precision to be inferior to a key press. We asked participants to balance accuracy and speed, and intentionally did not reveal their achieved accuracy after each trial (as done by others \cite{Chen1988,Hinckley1997}) to avoid a bias toward accuracy \cite{Hinckley1997}. In addition, to avoid participant response bias \cite{Dell2012}, we explicitly told them before the experiments that none of the techniques was developed by us.

\subsection{Apparatus}

\begin{figure}[tb]
  \centering
  \subfigure[Person interacting in the tangible condition.]{\label{fig:setup:a}\includegraphics[width=\columnwidth]{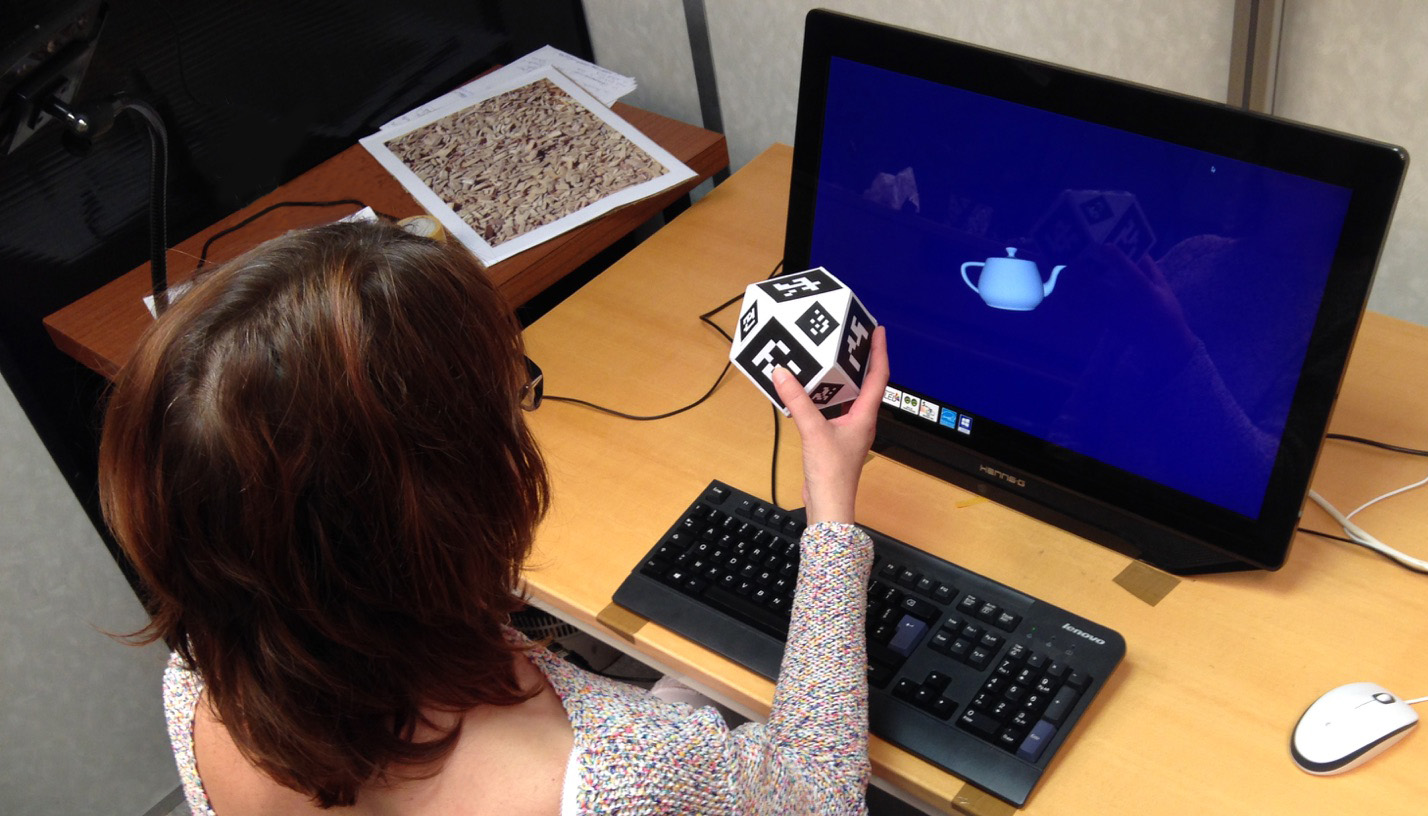}}\\[1ex]
  \subfigure[Tracking setup (cameras highlighted).]{\label{fig:setup:b}\includegraphics[width=\columnwidth]{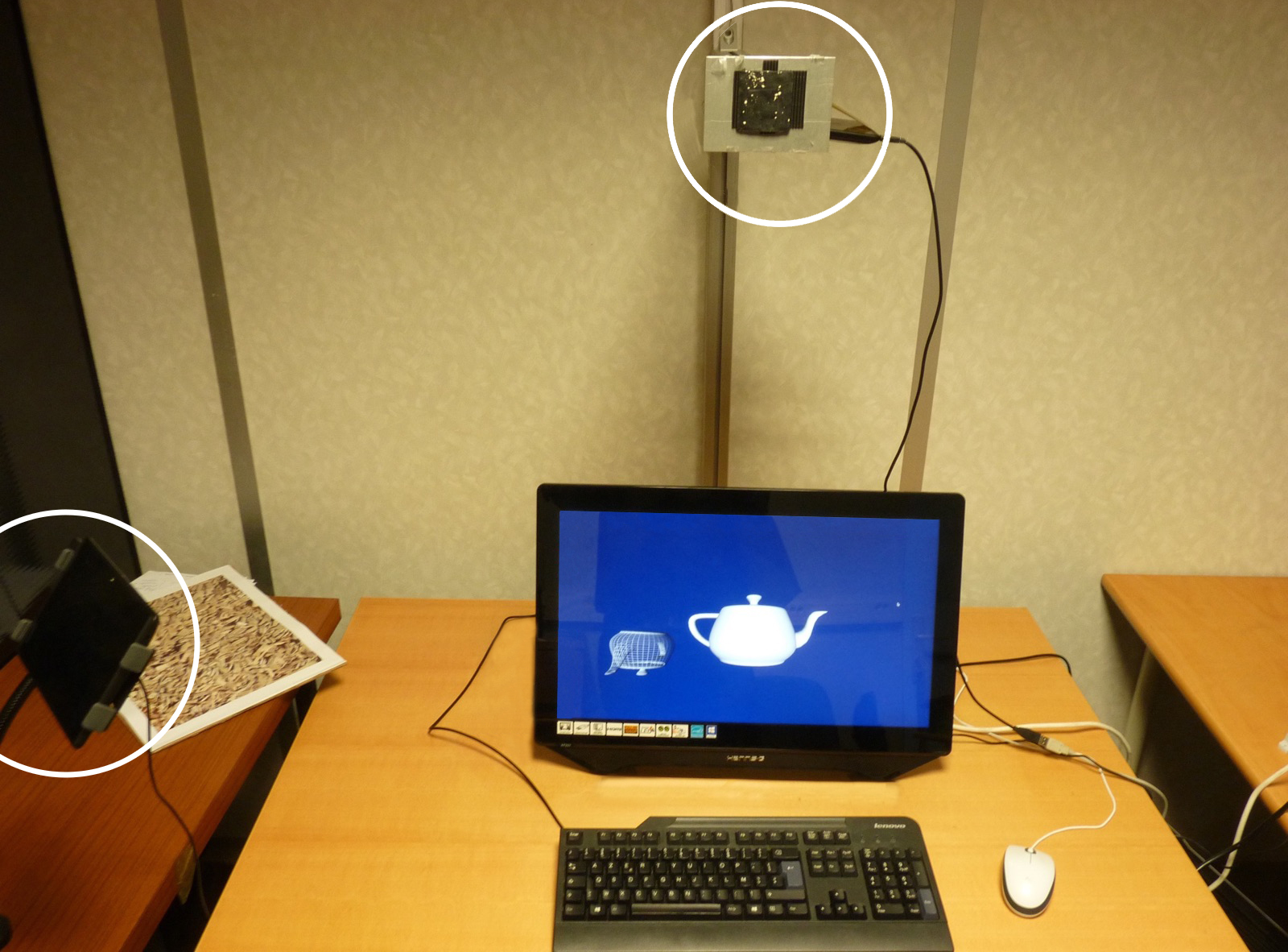}}
	\caption{Study setup.}
	\label{fig:setup}
\end{figure}

For all three input modalities we used the same touch-enabled 21'' LCD screen with a resolution of 1920\,\texttimes\,1080 pixels and a refresh-rate of 60\,Hz. Participants were asked to sit in front of the screen which was slightly tilted (approx.\ 15\textdegree) to provide a comfortable touch input setting (see \autoref{fig:setup}).
The mouse condition used a classical computer mouse: a Logitech m100 mouse at 1000 dpi with a polling rate of 125 Hz. The touch condition was evaluated using capacitive touch sensing built into the used screen. This touch sensor provided up to 10 points---captured via TUIO \cite{Kaltenbrunner:2005:TPT}.\footnote{See \href{http://www.tuio.org/}{http://www.tuio.org/}\,.}
The tangible condition was based on an optically tracked hand-held cardboard-based cuboctahedron (see \autoref{fig:setup:a}), each edge measuring 65\,mm. Markers on each face facilitated its 3D tracking with 6\,DOF. The tracking system comprised two cameras (see \autoref{fig:setup:b}), one located above to see both screen and tangible probe from above and one on the participants left side (approx.\  head level) so that the space in front of the screen can be seen. Both cameras together allowed us to avoid dead angles and ensured that participants could comfortably hold the cuboctahedron without blocking the camera's view.
\subsection{Interaction Mappings}

As much as possible, we chose established mappings for the evaluated input modalities as described next.

\textbf{Mouse+Keyboard.} Inspired by the mappings used by Blender,\footnote{See \href{https://www.blender.org/}{https://www.blender.org/}\,.} Autodesk MDT,\footnote{\textls[-5]{See \href{http://www.autodesk.fr/products/autocad-mechanical/overview}{http://www.autodesk.fr/products/autocad-mechanical/overview}}\hspace{.5pt}.} or Catia as well as software tools developed with VTK such as Paraview\footnote{See \href{http://www.vtk.org/}{http://www.vtk.org/} and \href{http://www.paraview.org/}{http://www.paraview.org/}\,.} we used the following mappings:
\begin{itemize}[itemsep=.25\lineskip,topsep=3\lineskip]
	\item left button: Virtual Trackball rotation for the $x$-/$y$-axes
	\item right button: translation along the $x$-/$y$-axes
	\item keyboard modifier + left button: rotation around the $z$-axis (leftward mouse motion = clockwise rotation)
	\item keyboard modifier + right button: translation along the $z$-axis
	\item the use of the scroll wheel was disabled since zooming needed to be inaccessible for the docking task
\end{itemize}
The Virtual Trackball (VT) technique relies on a virtual invisible sphere being created around the object to be rotated. While several rotation techniques have been implemented (for an overview see the surveys by Chen et al.\ \cite{Chen1988} and Bade et al.\ \cite{Bade:2005:UCM}), Bell's \cite{Bell:1988:LBT} VT and Shoemake's \cite{Shoemake1992} Arcball seem to be the ones most frequently used in available software tools. Yet, they are often seen as frustrating by users because they violate a number of principles for intuitive interaction \cite{Bade:2005:UCM}. Based on our pilot studies we decided to use an improved version of Bell's VT; one that respects the third principle mentioned by Bade et al.\ \cite{Bade:2005:UCM} and provides a transitive 3D rotation.

\label{sec:mappings:touch}

\begin{figure}
  \centering
	\includegraphics[width=\columnwidth]{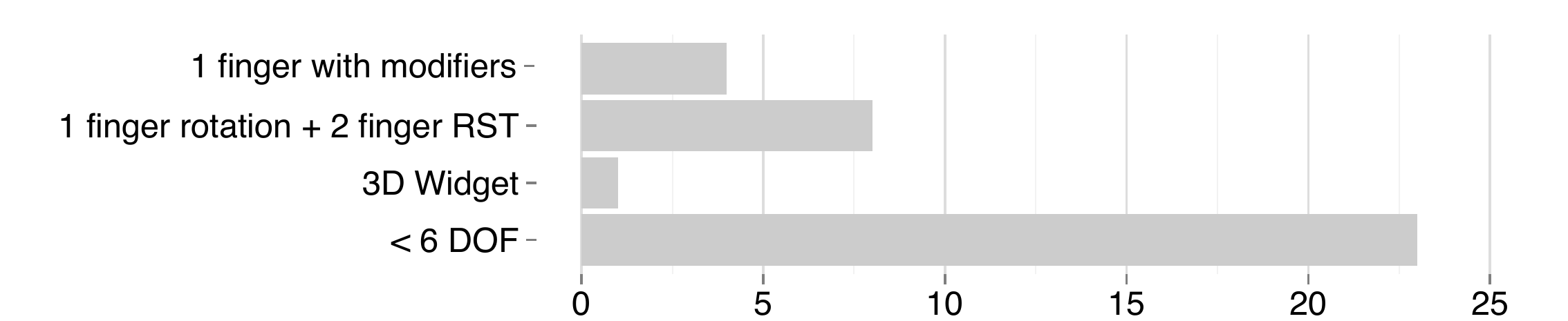}
	\caption{Touch mappings for 3D interaction in mobile apps.}
	\label{fig:touch-app-survey}
\end{figure}

\textbf{Touch Input.} In contrast to mouse+keyboard and tangible input, no single established standard or quasi-standard for touch-based interaction with 3D data exists. Based on our survey of 36 mobile applications on Android and iOS (see \autoref{fig:touch-app-survey}) we found that most interaction mappings do not provide the 6\,DOF we need. From those who do, most used the mapping that relies on either one or two fingers, with the latter providing rotation round the $z$-axis, uniform scaling, and translation along the $x$-/$y$-axes using pinching (RST). While studies have shown that it is possible to outperform the classical RST technique by separating the degrees of freedom \cite{Martinet2010}, we believe that the intuitiveness of the pinching mapping can be of advantage in our case, so we decided to use the following mappings:

\begin{itemize}[itemsep=.25\lineskip,topsep=3\lineskip]
	\item 1 finger motion: virtual trackball rotation for the $x$-/$y$-axes
	\item 2 fingers---RST
		\begin{itemize}[itemsep=.25\lineskip,topsep=0\lineskip]
			\item translation: translation along the $x$-/$y$-axes
			\item rotation: rotation around the $z$-axis
			\item pinching: translation along the $z$-axis (as also done by Hancock et al.\ \cite{Hancock2009})
		\end{itemize}
\end{itemize}

\textbf{Tangible Input.} As tangible input is not yet widely established outside academic research, we could not draw from established mappings in software tools. We thus decided to use the apparently intuitive isomorphic position control, \ie, a one-to-one mapping that moves and rotates the virtual object/scene similar to the motions of the tangible object in the real world. While such an interaction could be classified as a minimal TUI, it fulfills the four characteristics of TUIs as defined by Ullmer and Ishii \cite{Ullmer2000}---similar to other comparable tangible input devices in the literature \cite{Hinckley1994, Song2011}---and is thus well suited to evaluate the advantages and limitations of tangible interaction for 3D manipulation.

\subsection{Participants}

A set of 36 unpaid participants took part in our comparative study. Out of them, 10 were female and 26 were male; their ages ranged from 19 to 52 years (mean = 30.2, SD = 8.7; median = 26). Out of the 36 participants, 3 were left-handed, the remaining 33 right-handed. With respect to their expertise with 3D manipulation on a computer, 12 participants ranked themselves as skilled due to frequent use of video-games or 3D software, while 24 participants stated they had no significant prior experience. Furthermore, 22 of the participants had a university degree, while 14 had a high school degree. They all had either normal or corrected-to-normal vision.

\subsection{Procedure}

Participants were guided through the study by means of a study controller software that presented the different task blocks in turn. Before starting the trials of a new input modality, participants were introduced to the interaction technique. They were intentionally given minimal instruction in the use of each device, they were only informed that they could
\begin{itemize}[itemsep=.25\lineskip,topsep=3\lineskip]
\item	use the mouse's left and right buttons and the keyboard's shift key in the mouse+keyboard condition,
\item use multiple fingers on the tactile screen in front of them for the touch condition, and
\item use the tangible object for the tangible condition.
\end{itemize}
Further, the space in which the tangible object could be used was pointed out because participants had to keep within the field of vision of the cameras. An evaluator was present to answer potential questions during the experiment as well as take notes about the usage of each of the three input modalities.

Throughout the study, we asked participants to fill in several questionnaires. A first questionnaire captured their demographics as well as their level of fatigue before the experiment. After each condition, participants filled a questionnaire to assess their workload and fatigue level. For the former we used NASA's Task Load Index\footnote{See~\href{http://humansystems.arc.nasa.gov/groups/tlx/downloads/TLXScale.pdf}{\textls[-22]{http://humansystems.arc.nasa.gov\discretionary{/}{}{/}groups\discretionary{/}{}{/}tlx\discretionary{/}{}{/}downloads\discretionary{/}{}{/}TLXScale.pdf}}.}, the latter was based on Shaw's approach \cite{Shaw1998}. A final questionnaire assessed the subjective ratings for the different techniques. To confirm this last self-assessment, we informed participants that they would have to do a final set of 15 docking tasks, for which they could pick their favorite technique. Only after they had voiced their choice, we informed them that, in fact, the study was over and that the last question was only used to understand their true preferences. We finally asked whether, if given the free choice, they would have carried the additional batch of 15 tasks---to better understand people's eagerness to interact with the chosen technique.

\subsection{Variables}
With this comparative study we thus analyze one independent variable---the interaction modality---and five dependent variables---completion time, precision, fatigue, workload, and preferences. We decided to take two different types of accuracy into account: the Euclidean distance to the target in 3D space as well as the rotational difference (in degrees) to the target.

\subsection{Hypotheses}

Based on our previous experience with the three input modalities, we hypothesized that:\marginpar{\small should we address the comment on the hypotheses?}
\newlength{\desclabelwidth}
\settowidth{\desclabelwidth}{\textbf{H1}}
\addtolength{\desclabelwidth}{\labelsep}
\begin{description}[itemsep=.25\lineskip,topsep=3\lineskip,leftmargin=\desclabelwidth]
\item[H1] The time spent on trials would be shorter in the tangible condition than in the touch condition due to the inherent and fully integrated \cite{Jacob1994} structure. Touch-based interaction would also be faster than mouse-based input due to its higher directness and partially integrated structure.
\item[H2] The precision for both the rotation and the Euclidean distance to the target would be better for the mouse than the multi-touch condition due to the better support of the hand when using a mouse. The precision of the touch input, in turn, would be better than the tangible condition due to the lack of support for the hand when using tangibles.
\item[H3] The workload for the tangible condition would be low overall due to its intuitive mapping and fast interaction times---yet the need to have to hold the object and fine-position it would have a negative impact. The higher mental demand necessary to understand the mapping of touch and mouse interaction balanced by the reduced physical demand of these techniques would produce a slightly higher workload than for the tangible.

\item[H4] The resulting fatigue would be highest for the tangible due to having to hold the physical object at all times, lower for the touch input due to the added rest on the touch surface, and minimal for the mouse due to the arm being comfortably placed on the table.
\item[H5] People prefer both tangible input and touch input over mouse input, but for different reasons: touch for its ``intuitive'' mappings and reasonable precision, tangible because it benefits from the similarity to real-world interaction (but lacks a bit of precision). Mouse-based input is not preferred because it forces the separation of input DOF, while the others provide means of controlling several DOF in an integrated fashion.\end{description}

\section{Results}

We collected a total of 1620 docking trials from 36 participants, \ie, 540 trials for each input modality. To compare the three conditions, we measured the task completion times as well as an accuracy score for each condition and each participant based on their results in each of the trials for a given condition. 

While data from HCI experiment is traditionally analyzed by applying null-hypothesis significance testing (NHST), this form of analysis of experimental data has come under increasing criticism within the statistics \cite{Cumming2014} and HCI communities \cite{Dragicevic:2016:FSC,Dragicevic:2014:RAH}. We thus report our results using estimation techniques with effect sizes\footnote{The term \emph{effect size} here simply refers to the different means we measured. We do not refer to standardized effect sizes \cite{Coe2002}, as their report is not always recommended \cite{Baguley2009}, but rather to simple effect size.} and confidence intervals (instead of $p$-value statistics), consistent with recent APA recommendations \cite{VandenBos2009}.

\subsection{Task Completion Time}

\begin{figure}[tb]
 	\centering
  \subfigure[Completion times for each technique in seconds.]{\label{TimeFigure:a}\includegraphics[width=\columnwidth]{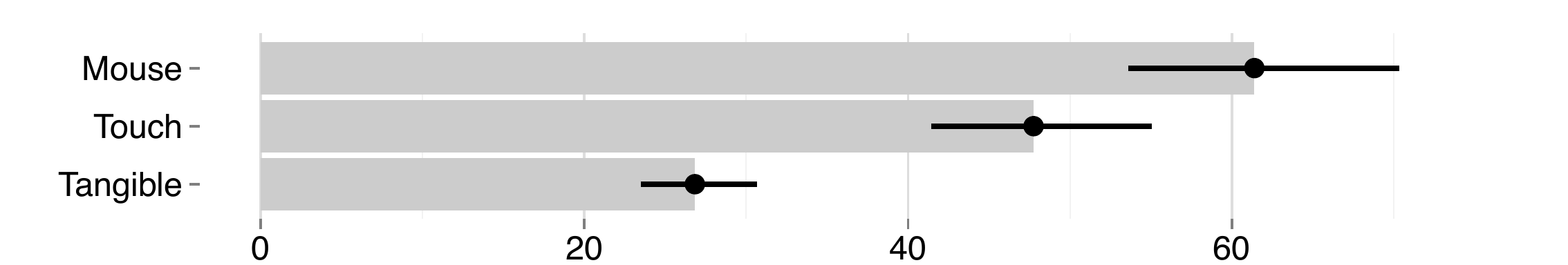}}\\
  \subfigure[Pairwise comparison ratios. Results of the left-side technique are divided by results on the right side. A ratio of 1 means similar performances.]{\label{TimeFigure:b}\includegraphics[width=\columnwidth]{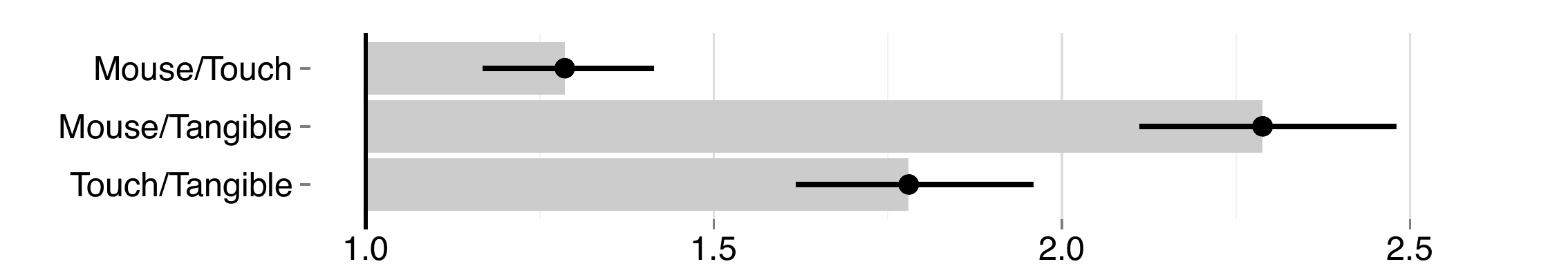}}
	\caption{Task completion times. Error bars are 95\% confidence intervals.}
	\label{TimeFigure}
\end{figure}

For the analysis, we log-transformed all time measurements to correct for positive skewness and present our results anti-logged, as it is standard in such cases \cite{Sauro2010}. Consequently, we arrive at geometric means. They dampen the effect of potential extreme trial completion times, which otherwise could have biased an arithmetic mean.

We present the completion time results in \autoref{TimeFigure:a}. It shows that it took participants 61\,s to complete the task in the mouse condition, 47\,s in the touch condition, and 26\,s in the tangible condition. While the confidence intervals reveal a difference in favor of the tangible condition over the mouse and touch conditions, they do not allow us to say anything more with confidence. We thus computed a pairwise comparison between the different conditions, see \autoref{TimeFigure:b}. The differences in these pairwise comparisons were also anti-logged and thus present ratios between each of the geometric means. These ratios all being clearly $\neq 1$ allows us to interpret the time differences of completing the task. \autoref{TimeFigure:b} shows that there is strong evidence for the tangible condition to clearly outperform the mouse condition, the tangible condition being more than twice as fast as the mouse condition. The difference between the tangible condition and the touch condition is also quite strong: the tangible condition is almost twice as fast as the touch condition. The difference between mouse and touch is not as strong; yet, the touch condition can still be considered faster than the mouse condition.

\begin{figure}[tb]
 	\centering
  \subfigure[Mouse condition.]{\label{fig:learning:a}\includegraphics[width=\columnwidth]{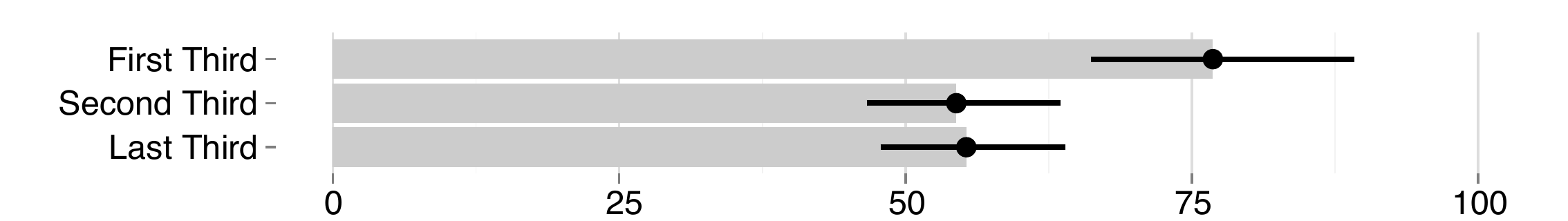}}\\[1ex]
  \subfigure[Touch condition.]{\label{fig:learning:b}\includegraphics[width=\columnwidth]{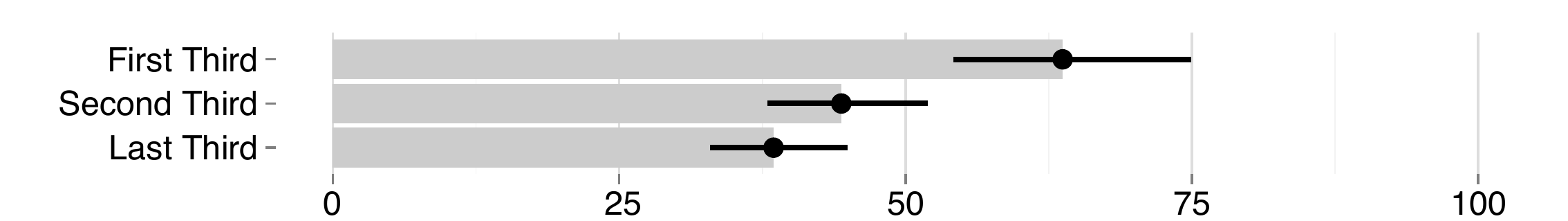}}\\[1ex]
  \subfigure[Tangible condition.]{\label{fig:learning:c}\includegraphics[width=\columnwidth]{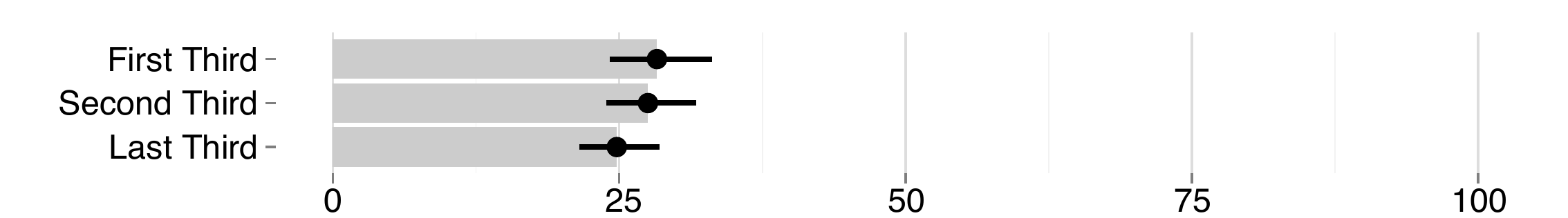}}
	\caption{Task completion times in seconds. Error bars are 95\% CIs.}
	\label{fig:learning}
\end{figure}

We also checked for learning effects. As shown in \autoref{fig:learning}, learning occurred for the mouse and touch conditions, at least when comparing the first block with the third block. This is most evident in the mouse condition (\autoref{fig:learning:a}) due to the lack of intuitive mapping for mouse-based interfaces. The touch condition similarly shows clear learning effects (\autoref{fig:learning:b}) as there is also a mapping involved that is not necessarily intuitive, despite the directness of this input modality. In the tangible condition (\autoref{fig:learning:c}) there is no evidence for learning. This may be due to the intuitiveness of the one-to-one mapping. Nevertheless, we did not find behavior different from that reported above when only analyzing the last two thirds or even the last third of the trials of the participants for the different conditions. 
\subsection{Precision}

\begin{figure}[tb]
	\centering
	\subfigure[Euclidean distances to the target for each technique in space units.]{\label{DistanceFigure:a}\includegraphics[width=\columnwidth]{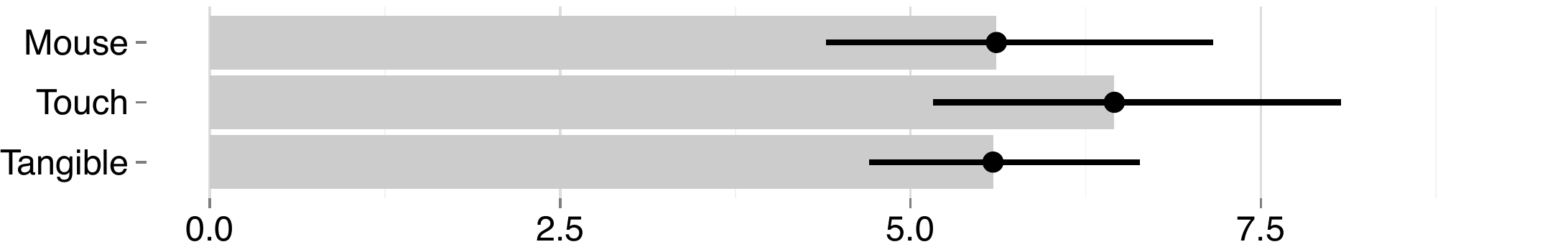}}\\[1ex]
	\subfigure[Pairwise comparison ratios for the Euclidean distance.]{\label{DistanceFigure:b}\includegraphics[width=\columnwidth]{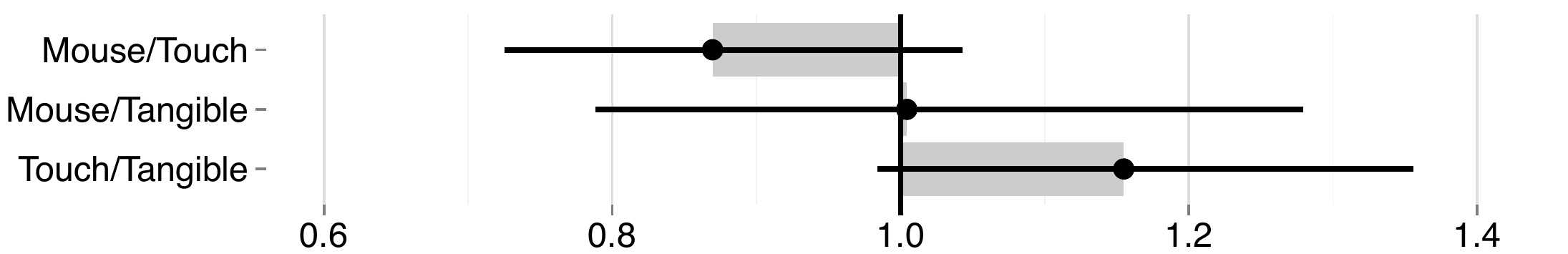}}
	\caption{Euclidean distances. Error bars are 95\% CIs.}
	\label{DistanceFigure}
\end{figure}

An inspection of Q-Q plots on the Euclidean and angular distance showed that the data did not follow a normal distribution but instead approximately followed a log-normal distribution. Like the time data, we also log-transformed both precision measurements for the analysis and, for better interpretation, present the results anti-logged.

\textbf{Euclidean Distance.} We report the Euclidean distances to the target in \autoref{DistanceFigure:a}. The figure shows that all three techniques lead to similar precisions, with means of 5\,mm for the mouse condition and the tangible condition, and 6\,mm for the tactile condition. Pairwise comparison between the conditions (\autoref{DistanceFigure:b}) suggest that the tangible and the mouse input may have a slight advantage over touch interaction, while both mouse and tangible inputs are very similar in precision to each other for our chosen task.

\begin{figure}[tb]
	\centering
	\subfigure[Rotational distances for each technique in degree.]{\label{RotationFigure:a}\includegraphics[width=\columnwidth]{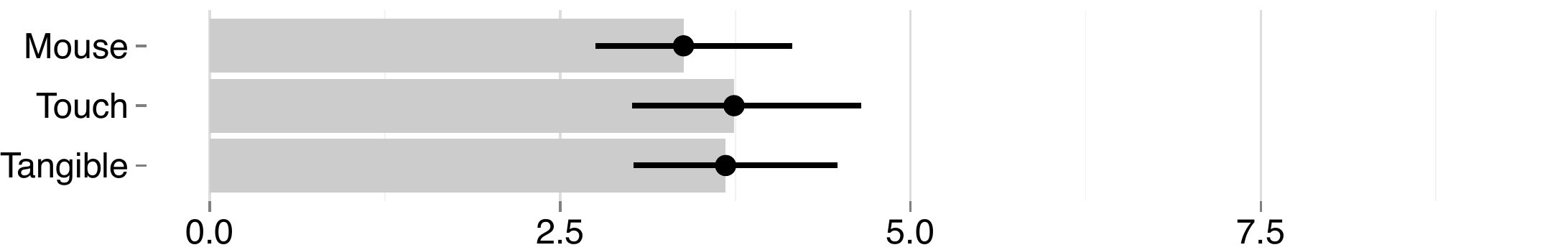}}\\[1ex]
	\subfigure[Pairwise comparison ratios for the rotational distances.]{\label{RotationFigure:b}\includegraphics[width=\columnwidth]{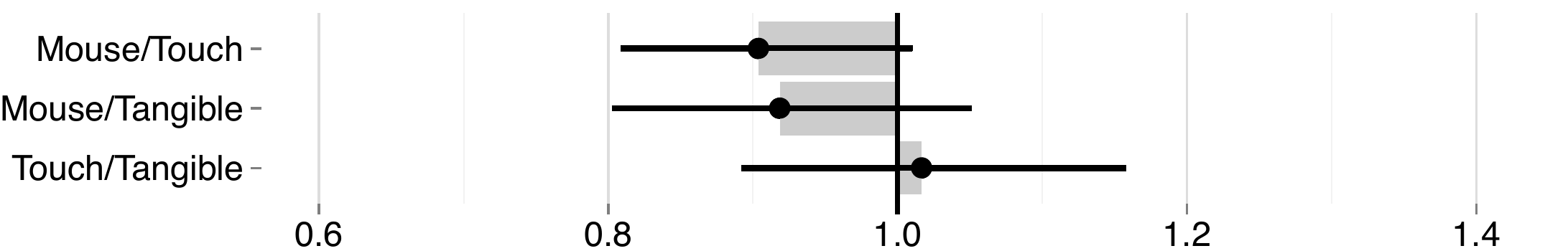}}
	\caption{Rotational distances. Error bars are 95\% CIs.}
	\label{RotationFigure}
\end{figure}

\textbf{Rotational Distance.} We report the rotational distance to the target in \autoref{RotationFigure:a}. The results are 3.4\textdegree\ for the mouse condition, 3.7\textdegree\ for the touch condition and the tangible condition. The pairwise comparison between the conditions is shown in \autoref{RotationFigure:b}. Similar to the Euclidean distance, these comparisons indicate that the three techniques are similar. There is weak evidence that the mouse may yield slightly more rotationally-precise results than touch and tangible. However we did not find evidence for a performance difference between touch and tangible for the rotation.

Having analyzed both types of precision, we did not find evidence for a large difference in precision between the different input modalities. This result did not change if we---to account for learning effects---only analyzed the latter 2/3 or even the last 1/3 of the trials of each participant in the different conditions. 

\subsection{Measuring Workload}

When collecting workload measurements by means of NASA's TLX we noticed that the pilot-study participants were often confused by its second part---weighing each one of the different sub-aspects (\ie, mental demand, physical demand, temporal demand, performance evaluation, effort, and frustration) for the task they were asked to accomplish. To avoid the seemingly random choices lead to inconclusive or even incorrect results we decided not to consider this second part of the TLX to be left with what is called a \textit{Raw TLX} (RTLX). According to Hart's \cite{Hart2006} survey, the RTLX may be equally well suited as the regular TLX. We thus compute the workload for each task as the average of the RTLX ratings by participants.

\begin{figure}[tb]
	\centering
	\includegraphics[scale=0.33]{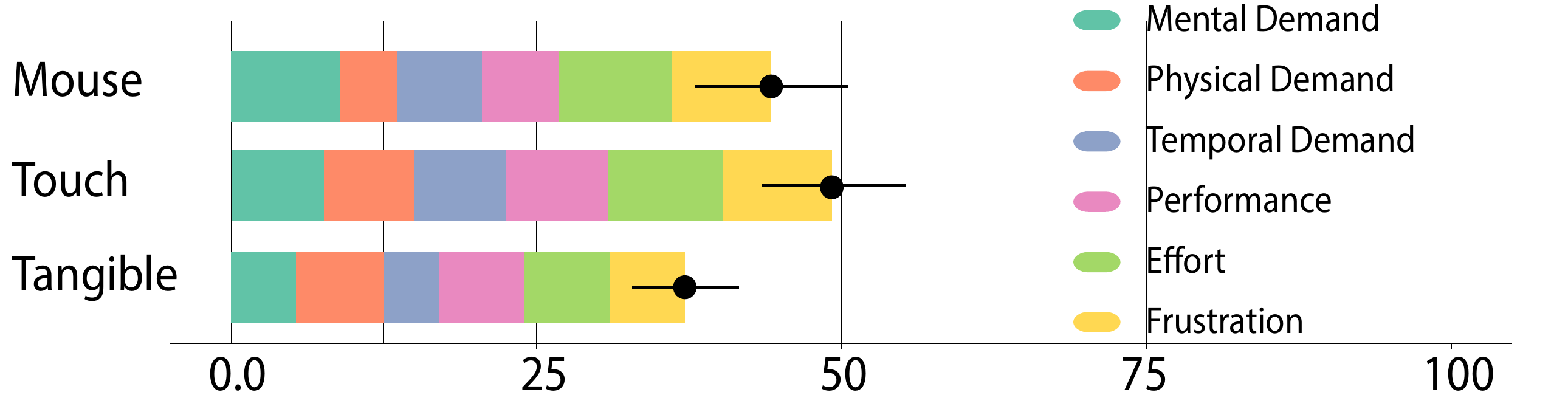}
	\caption{Total workload in overall NASA TLX units ($\in [0, 100]$). Error bars are 95\% CIs for the total workload ratings.}
	\label{FigureWorkload}
\end{figure}

The results of this analysis are shown in \autoref{FigureWorkload}. Here, we show the total workload for each of the conditions as well as the specific sub-aspects rated by participants. The non-overlapping confidence intervals between the touch and the tangible condition show that there the tangible condition requires a lower workload than the touch condition, yet for differences between the tangible and the mouse condition and even more so for differences between the mouse and the touch condition there is much less evidence. 

\begin{figure}[t!]
	\centering
	\subfigure[Mental demand.]{\label{subaspects:a}\includegraphics[width=\columnwidth]{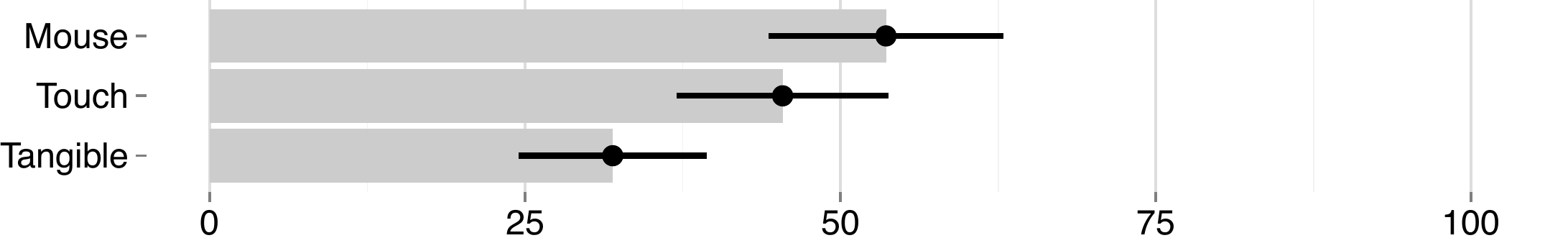}}\\[1ex]
	\subfigure[Physical demand.]{\label{subaspects:b}\includegraphics[width=\columnwidth]{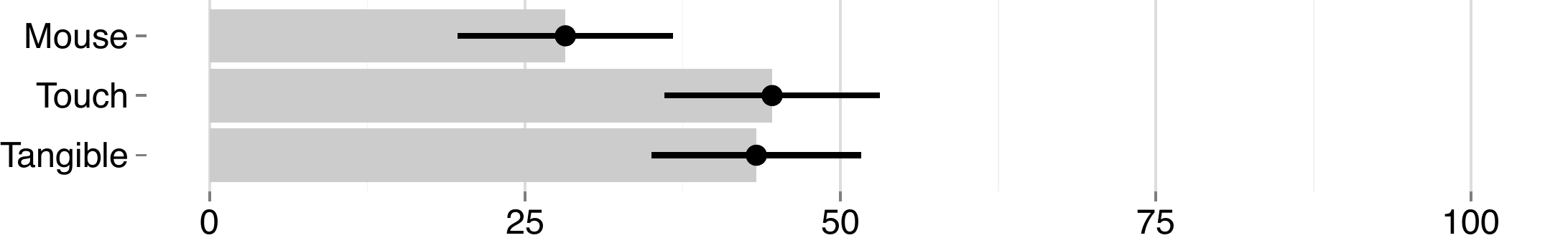}}\\[1ex]
	\subfigure[Temporal demand.]{\label{subaspects:c}\includegraphics[width=\columnwidth]{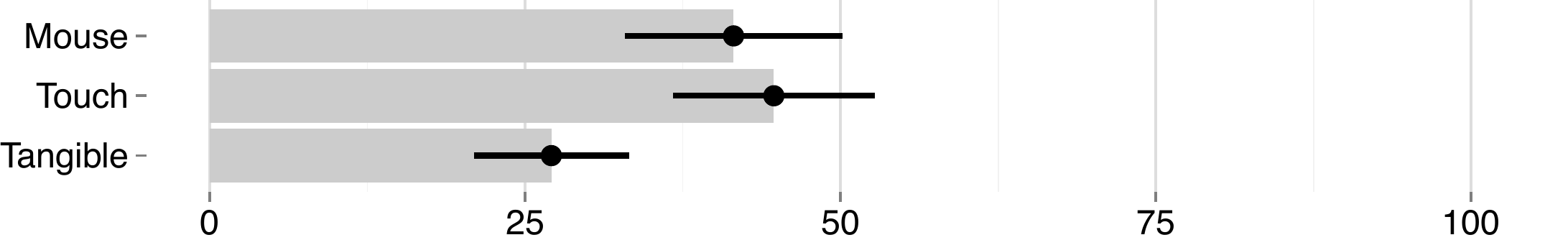}}\\[1ex]
	\subfigure[Performance (low score means good performance evaluation).]{\label{subaspects:d}\includegraphics[width=\columnwidth]{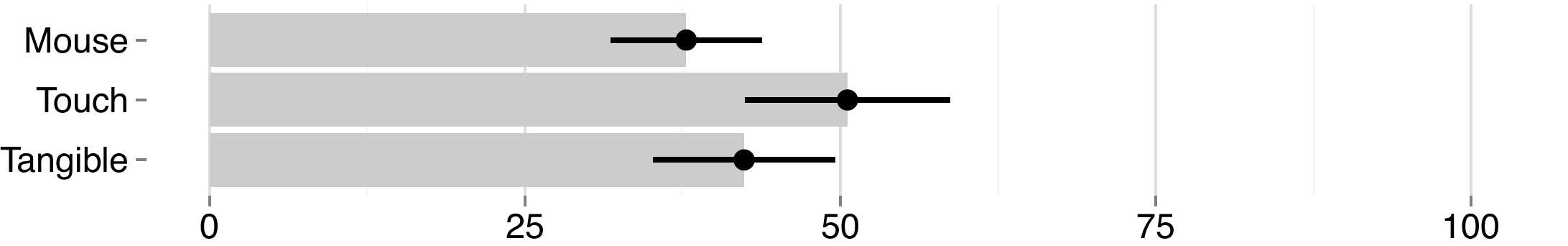}}\\[1ex]
	\subfigure[Effort.]{\label{subaspects:e}\includegraphics[width=\columnwidth]{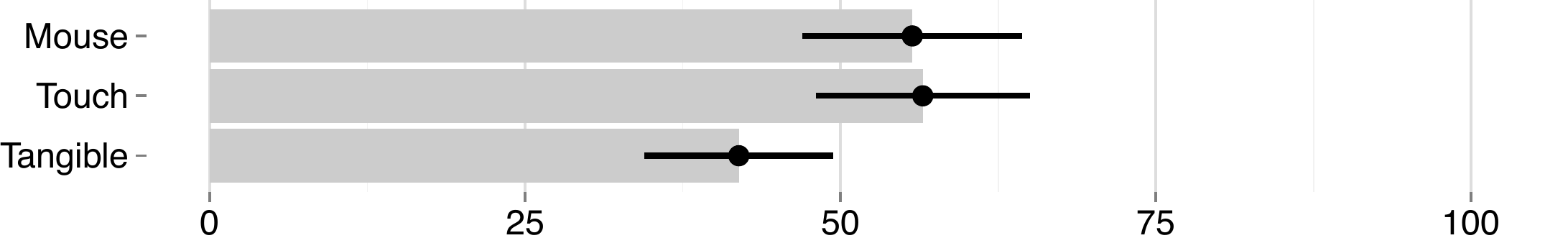}}\\[1ex]
	\subfigure[Frustration.]{\label{subaspects:f}\includegraphics[width=\columnwidth]{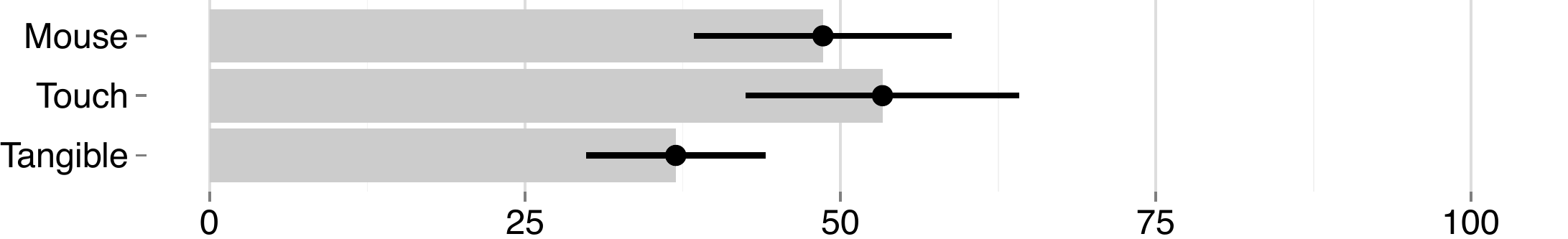}}
	\caption{Workload sub-aspects, detailed analysis of the data in \autoref{FigureWorkload} in individual NASA TLX units ($\in [0, 100]$). Error bars are 95\% CIs.}
	\label{subaspects}
\end{figure}

The individual sub-aspects of the workload differs somewhat between the different conditions, but we did not observe many striking differences between the three input modalities. \autoref{subaspects} shows a detailed analysis of the differences of the sub-aspects. We can observe that there are only clear differences in the rating of mental demand between the mouse and tangible condition (\autoref{subaspects:a}), for the physical demand between the mouse and the other two (\autoref{subaspects:b}), as well as for the temporal demand between touch and tangible condition (\autoref{subaspects:c}). The other comparisons between conditions for the sub-aspects only show gradual differences (also evident in the respective lengths of the colored patches in \autoref{FigureWorkload}). Yet, we can observe a slight advantage of mouse over touch for performance evaluation (\autoref{subaspects:d}), a small advantage of tangible over the other two for effort (\autoref{subaspects:e}), as well as a lower frustration in the tangible condition (\autoref{subaspects:f}). The difference in temporal demand between mouse and tangible (\autoref{subaspects:c}) matches the differences observed in overall interaction times between them (\autoref{TimeFigure}). In contrast, there was no difference between the mouse and touch condition even though we observed a clear difference in the completion time between them.

\subsection{Measuring Fatigue}

\begin{figure}[t!]
	\centering
	\includegraphics[scale=0.33]{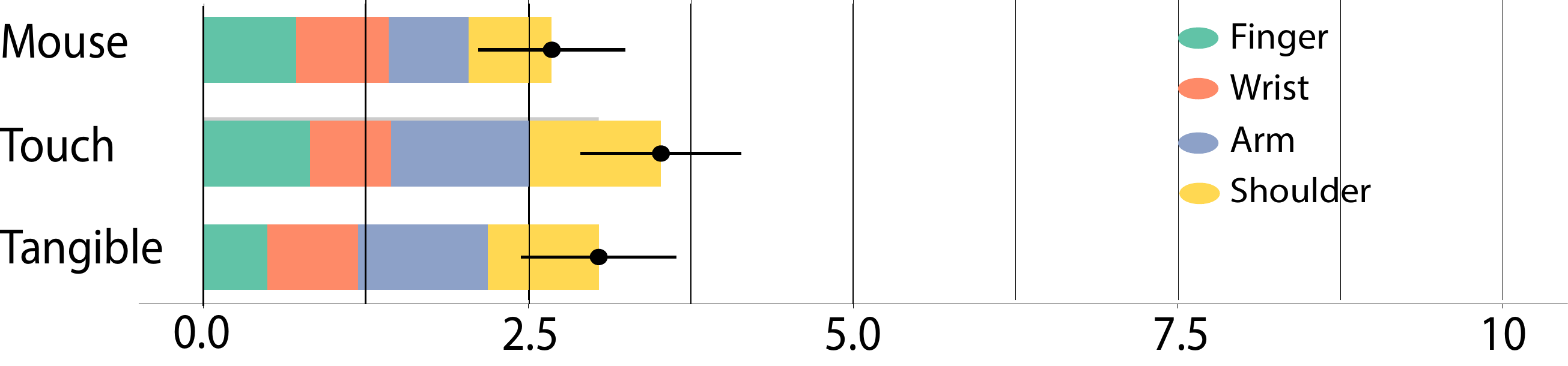}
	\caption{Total fatigue on a scale from 0 to 10. Error bars are 95\% CIs for the total fatigue ratings.}
	\label{FigureTotalFatigue}
\end{figure}

Similar to the analysis of the workload we present the analysis of the fatigue measurement in \autoref{FigureTotalFatigue}. Interestingly, none of the conditions exhibits a particularly high level of fatigue with the means all being lower than 4 on the scale of 0 to 10. While the mean of our measurements is highest for the touch condition, based on the confidence intervals there is no evidence that there would be an important difference between any of the conditions.

\subsection{Measuring Preferences}

\begin{figure}
	\centering
	\includegraphics[width=.65\columnwidth]{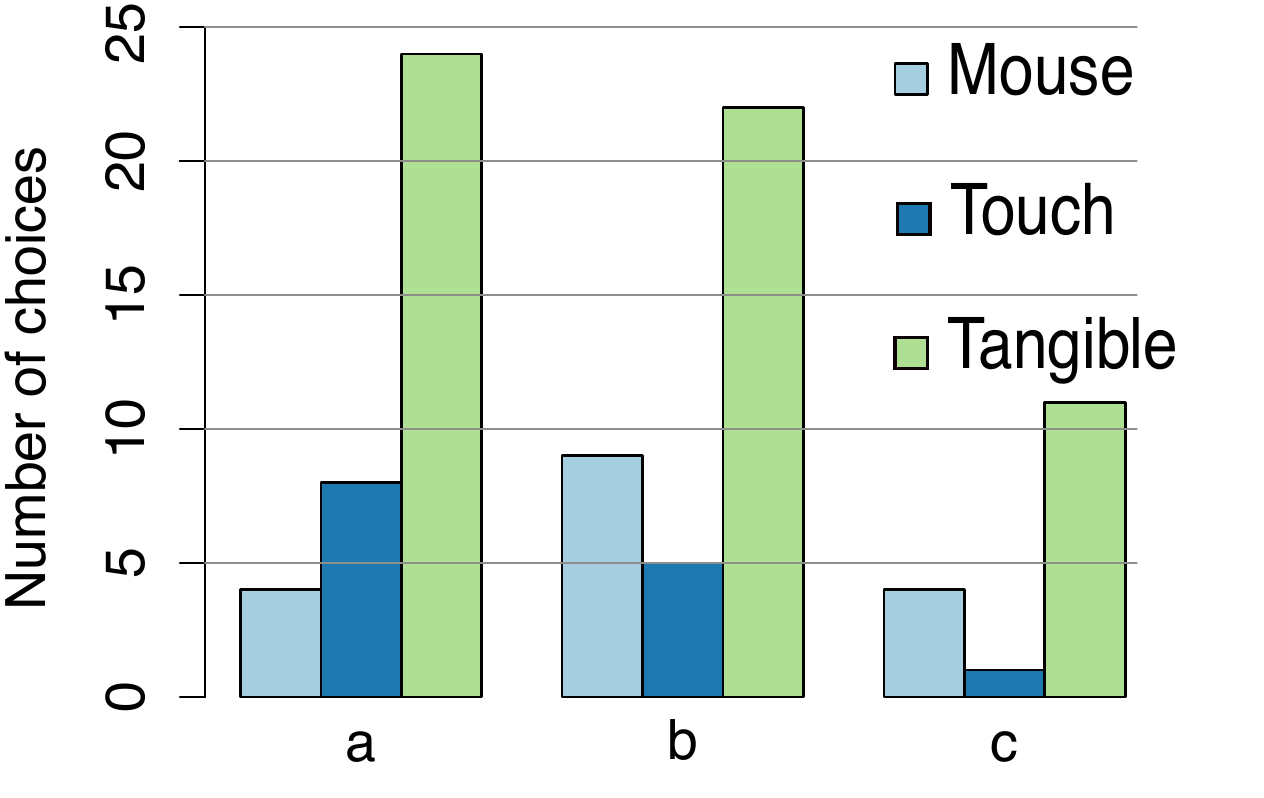}\vspace{-1ex}
	\caption{Participant preferences: (a) self-reported preferred technique, (b) technique chosen for the additional (but hypothetical) set of 15 trials, and (c) technique chosen by those participants who would have voluntarily stayed to complete the additional set of 15 trials.}
	\label{Preferences}
\end{figure}

We finally asked for participants' preferences. In addition to a normal preference rating---and as described in the previous section---we also asked participants which technique they would choose if faced with another set of 15 trials, and if they would want to actually stay for these additional 15 trials. \autoref{Preferences} reports these self-ratings.

Interestingly, the tangible condition was chosen most often for the stated preference (24\,\texttimes). Among those, however, 5 participants hesitated between touch and tangible, all ultimately picking the tangible as their favorite. The remaining 12 participants stated that they preferred touch over mouse (touch: 8\,\texttimes; mouse: 4\,\texttimes). When faced with an additional set of trials, a majority still preferred the tangible condition (22\,\texttimes). The touch vs.\ mouse preference, however, changed with the mouse now being rated higher than the touch (touch: 5\,\texttimes; mouse: 9\,\texttimes). Of the 16 participants who decided to do the tasks again ((c) in \autoref{Preferences}), 11 picked the tangible condition as their preference, 4 picked the mouse, and 1 picked touch.

\section{Qualitative Observations}

In addition to the quantitative analysis based on the captured data we also provide a summary of qualitative, observational data that was captured by the experimenter during the study.

\subsection{Touch}

We observed that many of the participants (20) had difficulties with using the two-finger RST interaction technique.  Our impression was that these difficulties arose from the RST technique integrating rotation and scaling into a single interaction, as opposed to wanting to only affect a single DOF at a time. For example, participants had trouble moving the two touching fingers without modifying their distance to each other, resulting in unwanted zoom-in or zoom-out actions while translating the object. Eleven participants actually stated that they would have preferred a mapping that would allow them to translate the object along the $x$-/$y$-axes without affecting its distance or its orientation around the $z$-axis.

We also observed that 20 participants used fingers from different hands for the pinch interaction, while 16 used two fingers from the same hand for the same interaction. This important difference in providing the input for the same type of interaction mapping likely had a large impact on people's precision and speed in completing the tasks as well as their preference. In both cases, however, participants reported the touch-based interaction and the corresponding interaction mapping to be ``intuitive'' and to be ``more natural than the mouse'' condition---13 participants made such statements when asked to assess the different conditions. Of them, 8 specifically praised the touch input for its perceived precision, while 5 reported that they thought they were faster with it than with the mouse condition. Two participants stated that they felt in control of the data they were manipulating, mirroring previous statements in other studies \cite{Watson:2013:DTE,Yu:2010:FDT}. Three participants, however, stated that they resented the fact that their action of removing their fingers from the touch screen led to little or even big transformations being issued inadvertently---the \textit{exit error} previously discussed by Tuddenham et al.\ \cite{Tuddenham2010}.

\subsection{Tangible}

The observation of participants interacting in the tangible condition showed that most of them were indeed not familiar with this type of input and manipulated the tangible object in interesting ways. For example, for rotations larger than 90\textdegree, 29 participants used two hands, while seven used only a single hand. Also, we observed that 12 participants completed the docking task using the tangible object by sequentially manipulating the different types of transformations. In contrast to an integrated interaction used by the other participants, they first translated the object along the $x$-/$y$-axes, then rotated it to match the orientation of the target, and finally translated it again to account for the $z$-translation. It is unclear, however, if the 12 participants who did not take advantage of the integrated interaction did not do this due to being used to the classical separated interaction offered by traditional 3D user interfaces, due to being afraid of loosing the optical tracking, or due to not feeling comfortable with the DOF-integrated manipulation offered by the tangible interaction. Finally, it is also worth noting that 17 participants (\ie, about half of them) used their non-dominant hand to interact if the docking target happened to be on the non-dominant side of the participant.

\section{Discussion}

With our ultimate goal of better understanding the different input modalities that are available for spatial manipulation in the context of the exploration of 3D scientific data, we now discuss those aspects of our results that are most surprising and/or most relevant for our target application domain.

\subsection{Completion Time}

In line with our hypothesis H1, we found that the tangible interaction was faster than the touch input which, in turn, was faster than mouse control. The reason for this difference in completion times is likely the inherent and straightforward integration of DOF control in the tangible condition, whereas the touch and mouse condition needed the switch of interaction modes with all the negative implications arising from user- or even system-controlled interaction modes (\eg, \cite{Buxton:1986:CPD,Sellen:1992:PME}). In addition, the touch condition provides some type of direct manipulation and DOF integration (\ie, 4\,DOF in the RST mode), while the mouse only controls 2\,DOF at any given time and is also an indirect input device.

We conjecture that, despite the advantages of the RST mapping (\autoref{sec:mappings:touch}), participants encountered difficulties with the mapping that may have impacted their performance, in particular the completion time.
We also hypothesize that the tangible condition's fast completion time may be a reason for its high precision: an approximate docking could be achieved a lot faster than in the other conditions, giving the participants more time to fine-tune their docking.

\subsection{Precision}

We initially thought that the different input modalities provided different degrees of precision. A mouse has a high-dpi sensor and a well-rested grasp configuration, while touch relies on the finger as a rather blunt instrument which also has less support. The tangible condition, finally, needs optical tracking with the arm operating in empty space. Yet, surprisingly, our data does not provide evidence for any of the three techniques providing higher precision (disproving H2). However, many participants still reported that they \emph{perceived} that they had precise control over their actions in the mouse (22\,\texttimes) and touch conditions (8\,\texttimes). In the tangible condition, however,they felt that they had uncontrollable and involuntary hand movements. We believe that this \emph{perceived} level of precision should not be disregarded in a decision of which interaction device to use or to offer for tasks that require a high precision.

\subsection{Workload}
With our data, we cannot confirm our hypothesis H3, but the overall measurements show---for our task and participant group---the same tendency as argued in the hypothesis: The perceived workload for the tangible interaction is lower than for the touch condition slightly lower than for the mouse condition. We believe, however, that the touch input (as well as the mouse input) can be improved. We saw that many participants kept their arms in the air while interacting using touch which contributed to the workload. This issue could be improved upon using a better (touch-only) setup and a better interaction mapping. With respect to the latter, we noticed that many participants had problems with the sensitivity of the $z$-translation---caused by them starting the interaction with their fingers very close together as they are used to interact on smart phones and tablets. Touch interaction---even or in particular if it uses the same interaction mappings---may require people to re-learn some of their familiar interaction techniques as they transition from small to larger screens.

Similarly, we also observed some frustration in the tangible condition. Some participants who felt at ease with the tangible interaction tried to manipulate it fast with one or two hands. Our optical tracking system, however, was only good enough for slow to medium movements but could not follow relatively fast manipulations, leading to frustration for participants. Similarly, participants occasionally occluded both cameras of our tracking system, leading them to report frustration due to the interruption of the tracking---maybe even focusing on such issues when rating the frustration and not concentrating on other interaction issues.

\subsection{Fatigue}

Based on the fatigue measures obtained in our experiment, we cannot confirm our hypothesis H4. The study setup was created such that---to facilitate a fair comparison---there was both enough space for mouse-based and tangible input as well as an equivalent view on the screen for all conditions. 
This, however had an implication on the self-assessed fatigue values. Indeed, many participants did not rest their elbows in the touch condition potentially resulting in shoulder and arm fatigue that would probably not have been perceived on a touch-dedicated setup. Such a setup would have also reduced the physical demand of the workload for touch interaction. Nevertheless, the fatigue ratings for all techniques are quite similar to each other, so that at least the fatigue measurement seems to have little impact on the choice of interaction modality.

We would also like to emphasize that tangible interaction lacks the possibility to easily maintain the virtual object in a given position and orientation as people release it. This was reported by four participants when they were asked what they liked about each condition. We could thus conjecture that an extended use of the tangible could drastically impact the fatigue level if users are not offered the option to release the tangible object without causing exit errors.

\subsection{Preferences}

Our data clearly showed an overwhelming preference for the tangible interaction, thus contradicting our hypothesis H5. However, we believe that this result should be taken with a grain of salt. Our participants' preference for tangible is likely biased by them being used to mouse and touch-based interaction, while tangible input is new to the vast majority of them. Indeed, some of the participants who selected tangible input as their favorite explained that they would use this technique for the forced and free choice (\ie, (b) and (c) in \autoref{Preferences}) because they do not have the opportunity to ``play'' with this kind of technology at home, while they have an easy access to touch screens and mice: This clearly made a difference at least for 5 out of the 11 participants who picked the tangible option for the last preference choice (\ie, (c) in \autoref{Preferences}).

\subsection{Realistic Application Scenarios}

While our study scenario and task were chosen to be representative of generic 3D interaction as needed for visual data exploration, for realistic scenarios we likely face different interaction requirements. For example, we envision that longer interaction periods will be needed with different types of tasks and more complex interaction techniques. The longer interaction periods will have an effect on fatigue and workload, in particular for tangible interaction and for touch input. Realistic tasks, moreover, require more than 6\,DOF interaction: uniform or non-uniform scaling are needed as well as interactions constrained to specific DOF should at least be included. In addition, many other interaction modalities are needed for practical applications such as cutting plane interaction, parameter specification, view selection, data selection, etc.\ (\eg, \cite{Coffey:2012:ISW,Keefe:2013:RSV,Yu:2010:FDT}). All these are likely to favor mouse- and touch-based input, as tangible interaction will likely be more difficult to use for generic interaction unless multiple tangible input devices are used. Tangible input, however, may have some benefits for specialized input (\eg, \cite{Jackson:2013:LTI,Sultanum:2011:PSP}), while touch input may be better for integrated approaches (\eg, \cite{Coffey:2012:ISW,Klein:2012:DSD,Sultanum:2011:PSP}). A final aspect to consider for realistic application scenarios is that in these, unlike the participant population we tested, we would be faced with experts in 3D interaction as they carry out such tasks on an everyday basis. Even though the learning effects we saw did not affect the results of our study overall, we may see other preference ratings among domain experts after longer periods of use than the ones voiced by our participants.

\subsection{Summary of Limitations}

The discussion so far has, in fact, mentioned many of the limitations of this work already, so we only provide a brief summary here. Our study was limited by the need for a setup that would accommodate all three input modalities, while in practice dedicated setups better suited to a given modality would lead to better results. Moreover, practical applications will require more complex interaction scenarios, for which mouse and touch-based input are likely better suited than tangible interaction. In addition, the chosen participant population for a quantitative experiment such as this one is different for the ultimate target audience, and the novelty factor of tangible interaction also introduced a bias---in particular for the self-reported preferences. Another influence of the chosen participants is that we faced learning effects, that would disappear if the techniques would be used in practice for a longer time. Finally, the chosen mapping for, in particular, touch interaction may be successful in one type of application, but other applications and combinations with additional interface elements may require other mappings that may better be suited for visual exploration of 3D data. We believe that this mapping question should be the focus of future research.

\begin{table}[tb]
	\tabulinesep=2pt
  \centering
  \caption{\label{table:summary} Advantages and limitations of each input modality.}\vspace{.7\abovecaptionskip}
  \footnotesize
  \begin{tabu}[\columnwidth]{@{}%
	  l@{\hspace{5pt}}%
		X@{\hspace{5pt}}%
		X
		c@{}}
		\toprule
     & advantages & disadvantages \\
    \midrule
	\centering mouse &
	\vspace{-.85em}
    	\begin{itemize}[leftmargin=*, topsep=0pt, noitemsep,labelsep=0.2em]
	\item availability
	\item perceived precision
	\item familiarity
	\item DOF separation
	\item low physical fatigue
	\item moding for complex tasks
	\vspace{-1mm}
	\end{itemize}
	&
	\vspace{-.85em}	
    	\begin{itemize}[leftmargin=*, topsep=0pt, noitemsep,labelsep=0.2em]
	\item difficult mapping
	\item slowest interaction
	\item moding required
	\end{itemize}
\\
\midrule
	\centering touch &
	\vspace{-.85em}
    	\begin{itemize}[leftmargin=*, topsep=0pt, noitemsep,labelsep=0.2em]
	\item availability
	\item perceived precision
	\item increased directness
	\item faster than mouse
	\item easier mapping
	\item multiple mappings for complex tasks
	\vspace{-1mm}
	\end{itemize}
	&
	\vspace{-.85em}
    	\begin{itemize}[leftmargin=*, topsep=0pt, noitemsep,labelsep=0.2em]
	\item unclear suitability of given mappings
	\item slower than tangible
	\item physical fatigue
	\item exit error
	\vspace{-3mm}
	\end{itemize}
\\
\hline
	\centering tangible &
	\vspace{-.85em}
    	\begin{itemize}[leftmargin=*, topsep=0pt, noitemsep,labelsep=0.2em]
	\item fastest interaction
	\item intuitive mapping
	\item impression of control
	\item novelty factor
	\vspace{-1mm}
	\end{itemize}
	&
	\vspace{-.85em}
    	\begin{itemize}[leftmargin=*, topsep=0pt, noitemsep,labelsep=0.2em]
	\item complex tasks not supported
	\item relies on 3D tracking
	\item needs a separate object
	\item inflexible interaction mapping
	\item always on, extra moding needed to stop interacting
	\item exit error
	\vspace{-2mm}
	\end{itemize}
\\

		\bottomrule
    \end{tabu}
\end{table}

\section{Conclusion}

Despite these limitations, however, our study has provided valuable insights on the potential of the three input modalities---mouse, touch, and tangible---for the use in 3D interaction in general and, specifically, for the visual exploration of 3D data. In particular, we found that they are all equally well suited for precise 3D positioning tasks---contrary to what is generally assumed about touch and tangible as input modalities. Analysis of task completion time showed that tangible interaction was fastest, touch slower, and mouse slowest. However, we did observe learning effects that may play out for longer-term usage, even though our data still showed the same advantage for tangible interaction if only the last third of trials was examined. Moreover, we discussed several additional considerations that need to be taken into account when designing practical interaction scenarios that put the observed advantages of tangible interaction into perspective. Researchers can now build on our findings by knowing that there is not a single input modality that would be a clear favorite for controlling 3D data during visual exploration, but that all three have their respective advantages and disadvantages that should be considered and are summarized in \autoref{table:summary}.

Our findings also facilitates further studies that can now focus on other aspects of the different input modalities. In particular, the interaction mapping for touch input will remain a focus of future research. In addition, the issue of the exit error will have to be addressed for both touch and tangible inputs. The presence or the lack of spatial multiplexing of DOF control for touch (which some participants did not use despite this being possible) is another aspect that should be investigated. A closer investigation of people's use of dominant and non-dominant hands during interaction for both the tangible and the touch conditions also would be an interesting path to follow. Ultimately, however, we want to continue our examination of how to best create an interaction continuum that allows one to fluidly switch between different interaction scenarios and interaction environments---picking the best one for a given task or situation. This direction of work will be facilitated by the insights we gained with this study.


\bibliographystyle{abbrv-doi-doi-narrow}
\bibliography{MouseTouchTangible}
\end{document}